\documentstyle[prl,aps]{revtex}
\begin{document}

\title{Colliding black holes: The close limit}

\author{Richard H. Price}
\address{Department of Physics, University of Utah, Salt Lake City, UT
84112-1195}

\author{Jorge Pullin}
\address{Center for Gravitational Physics and Geometry, 
The Pennsylvania State University, University Park, PA 16802}
\date{February 22, 1994}

\maketitle

\begin{abstract}

The problem of the mutual attraction and joining of two black holes is
of importance as both a source of gravitational waves and as a testbed
of numerical relativity. If the holes start out close enough that they
are initially surrounded by a common horizon, the problem can be
viewed as a perturbation of a single black hole. We take initial data
due to Misner for close black holes, apply perturbation theory and
evolve the data with the Zerilli equation.  The computed gravitational
radiation agrees with and extends the results of full numerical
computations.

\end{abstract}


\vspace{-8.5cm} 
\begin{flushright}
\baselineskip=15pt
CGPG-94/2-3  \\
gr-qc/9402039\\
\end{flushright}
\vspace{7cm}

The collision of two black holes is, in principle, one of the most
efficient mechanisms for generation of gravitational waves. In view of
the fact that the LIGO and VIRGO \cite{Ab} detectors may be detecting
events in the coming years, the theoretical determination of possible
waveforms has become of great importance. The fact that data may be
well below the noise level of the detectors may require
pattern-matching techniques \cite{Cuetal} which require accurate
knowledge of radiation waveforms.

The problem of the gravitational radiation generated by colliding
black holes is not only of great importance to gravitational wave
astrophysics, it has also been one of the earliest applications of
numerical general relativity. Smarr and
Eppley\cite{SmEp,smarrsources}, more than 15 years ago, computed the
radiation waveforms for the axisymmetric problem of two holes,
starting from rest and falling into each other in a head-on collision.
The importance of this problem has motivated a recent reconsideration,
both numerical and analytical, by Anninos {\em et al.}\cite{Annetal}.
The numerical work is difficult, especially when the holes are
initially close together. In that case the radiation is dominated by
horizon processes. In addition, for initially close holes the
radiation generated is relatively small and the numerical errors in
its computation can be particularly troublesome.  The purpose of this
paper is to provide a method of computing the radiated power generated
when the holes start off close together. Our method is based on
perturbation theory and is considerably more economical than a full
numerical simulation. It can also be viewed as a benchmark against
which numerical codes can be checked.

For a full numerical computation of the problem, data used are those
for the two throats of a momentarily static ``wormhole,'' for which an
analytic solution was given by Misner\cite{Misner}. The initial data
has a parameter $\mu_0$ which can be adjusted so that the initial
conditions correspond to different values of $L/M$, where $L$ is the
initial separation of the throats, and $M$ is the mass of the
spacetime.

For values of $\mu_0$ corresponding to large
and moderate starting values of $L/M$ the motion of the
``particles,'' not their black hole nature, is crucial to the
generation of gravitational radiation. In this case the amount of
radiation emitted can be understood with a quasi-newtonian
approximation that starts with the known radiation for a point mass
falling into a hole\cite{Annetal}. For values above around $\mu_0
\approx2$, this quasi-newtonian approximation is in remarkably good
agreement with the results of numerical relativity. For smaller values
of $\mu_0$, however, the approximation seriously overestimates the
radiation.  When the throats start off at small separation
their ``internal'' structure cannot be ignored.

It is our purpose here to provide an understanding of the opposite
limit, the limit of small $\mu_0$ (and hence small $L/M$). Below we
give a simple and attractive analytic result valid in the
limit of small separation for the energy radiated. A rather complete
understanding of the phenomenon is then afforded by this result along
with the quasi-newtonian estimate, and the bridge between the two
given by the numerical relativity computations.  In addition, our
method gives predicted waveforms, and other features of the wave, that
can be useful for verifying the accuracy of the numerical relativity
computations.

It should be understood that our small-$\mu_0$ result does not have
the same robust connection with a simple physical picture as the
quasi-newtonian approximation has with the mutual in-fall of two holes
from large distance. In the large distance limits the details of the
choice of initial data to represent the individual holes is
unimportant; it is their particle-like motion toward each other that
dominates the radiation. For the initially close limit the Misner data
is one possible choice for momentarily static initial conditions, but
it is singled out by mathematical convenience, not by any claim that
it represents the most natural, instantaneously stationary, initial
distortion of the participating throats.

In terms of bispherical coordinates $\mu, \eta, \phi$, the Misner
initial data\cite{Misner} take the form
\begin{equation}                                              
ds^{2}_{\rm Misner}=a^{2}\varphi^{4}_{\rm
Misner}\left[d\mu^{2}+d\eta^{2} +\sin^2{\eta}\,d\phi^{2}\right]\,,
\end{equation}
where $a$
is a constant with the dimension of length, and 
\begin{equation}                                                
\varphi_{\rm
Misner}=\sum_{n=-\infty}^{n=-\infty}\frac{1}{\sqrt{\cosh(\mu+2n\mu_0
)-\cos\eta}}\,.
\end{equation}

We now change from bispherical to spherical coordinates
$R,\theta,\phi$ and introduce a Schwarzschild radial coordinate $r$
through $R=\left(r^{1/2}+\sqrt{r-2M}
\right)^{2}$ to arrive at a line element of the form
\begin{equation}                           
ds^{2}_{\rm
Misner}=F(r,\theta;\mu_{0})\left[\frac{dr^{2}}{1-2M/r}+r^{2}d\Omega^{2}
\right]\,,
\end{equation}
with $d\Omega^{2}=d\theta^{2}+\sin^{2}\theta\, d\phi^{2}$, and with
\begin{equation}
F=\left(1-\frac{M}{2R}\right)^{-4}\left(1+\frac{a}{R}\sum_{n=0}^{\infty}
\frac{1}{\,\sinh{n\mu_{0}}\sqrt{1+2a/R\coth{n\mu_{0}\cos\theta
+a^{2}/R^{2}\coth^2{n\mu_{0
}}}
}}
\right)^{4}\,.
\end{equation}
The square root in the summation has the form of the generating
function for the Legendre functions $P_{\ell}(\cos{\theta})$, so that $F$
can be expressed as
\begin{equation}
\label{Feq}
F=\left(
1+\frac{2}{1+M/2R}
\sum_{\ell=2,4\ldots}^{\infty}\kappa_{\ell}\left(\frac{M}{R}\right)^{\ell+1}
P_{\ell}(\cos{\theta})
\right)^{4}\,,
\end{equation}
with
\begin{equation}
\kappa_{\ell}\equiv
\frac{1}{\left[4\Sigma_{1}({\mu_0})\right]^{\ell+1}}\sum_{n=1}^{\infty}
\frac{(\coth{n\mu_0})^{\ell}}{\sinh{n\mu_0}}\,.
\end{equation}
Here, and below, we use the notation: 
\begin{equation}\label{sigmanotat}
\Sigma_{k}({\mu_0})
\equiv\sum_{1}^{\infty}(\sinh{n\mu_{0}})^{-k}\,.
\end{equation}

Since the Misner geometry satisfies the initial value equations of
general relativity, and is momentarily stationary, there exists a
coordinate choice $T,r,\theta,\phi$ such that the initial data
generates a 4-geometry at 
$T=0$ of the form $ds^2=-dT^{2}+ds^{2}_{\rm Misner}$,
and for which $\partial g_{\mu\nu}/\partial T=0$
at $T=0$. One can make a transformation such that the 
4-geometry  takes the form
\begin{equation}\label{SchForm}
ds^{2}=-\left(1-\frac{2M}{\tilde{r}}\right)dt^{2}+F(r,\theta;\mu_{0})\left[
\frac{d\tilde{r}^{2}}{1-2M/\tilde{r}}+\tilde{r}^{2}d\Omega^{2}
\right]+
{\cal O}(t^{2})\,.
\end{equation}
For a constant $t$ slice of the Schwarzschild geometry $F$ is unity.
The difference between unity and the Misner form of $F$
is the extent to which the geometry in (\ref{SchForm})
initially deviates from a spherically symmetric black hole.

We have therefore cast Misner's initial data in Schwarzschild
coordinates. We now need to explore the limit in which the two black
holes are ``close.'' When $\mu_{0}<1.36$ an apparent horizon at
$R\approx 2M$ surrounds both throats. As $\mu_{0}$ decreases further,
the value of $L/M$ decreases, and hence the ratio of $L$ to the
horizon radius decreases. As pointed out by Smarr the 
horizon, and the geometry outside it, should then be nearly spherical,
and it is only the geometry outside the horizon that influences the
radiation sent outward to infinity. Linearized perturbation theory
should therefore give a good  description of the
generation of radiation (though not of the highly non-spherical
geometry inside the horizon).

To give this picture a mathematical realization we note that the
coefficients in (\ref{Feq}) then have the small-$\mu_{0}$ limit
\begin{equation}\label{kappapprox}
\kappa_{\ell}\approx\frac{\zeta(\ell+1)}{(4|\ln{\mu_0}|)^{\ell+1}}\,.
\end{equation}
In the spacetime corresponding to (\ref{Feq}) we can therefore
consider $\epsilon\equiv1/|\ln{\mu_0}|$ to be an expansion parameter.
If in (\ref{Feq}) we keep only the leading term in $\epsilon$
for each $\ell$ we get
\begin{equation}\label{lowest}                         
ds^{2}_{\rm Misner}\approx
\left[
1+\frac{8}{1+M/2R}\sum_{\ell=2,4,\ldots}^{\infty}
\kappa_{\ell}\left(\frac{M}{R}\right)^{\ell+1}
P_{\ell}(\cos{\theta})
\right]
\left[\frac{dr^{2}}{1-2M/r}+r^{2}d\Omega^{2}
\right]\,.
\end{equation}
We now argue that each multipole term can be treated individually by
linearized perturbation theory.  This is clearly true for the $\ell=2$
case.  To lowest order in $\epsilon$, that is to order $\epsilon^3$,
the field equations contain only terms linear in the perturbation
(i.e., linear in $1-F$), and those terms are pure $\ell=2$. Next
consider, for example, expansion to order $\epsilon^7$.  This would
contain $\ell=6$ terms linearly, but would also contain nonlinear
terms, e.g., from the square of the $\ell=2$ term (with higher order
corrections to (\ref{kappapprox}) included). But the nonlinear
contributions can have no $\ell=6$ part. (To get an $\ell=6$ term
requires a cube of an $\ell=2$ term, or a product of $\ell=2$ and
$\ell=4$, both of which are higher order in $\epsilon$.) This shows
that the $\ell=6$ part of the vacuum Einstein equations is linear in
$1-F$, and hence can be treated by linear perturbation theory. This
argument easily generalizes to arbitrary $\ell$.

Though each of the $\ell$-poles in (\ref{lowest}) satisfies the
linearized source-less Einstein equations, they are not all the same
order in $\epsilon$. We take up here only the dominant term at small
initial separation, the pure quadrupole $\ell=2$ term.  To treat this
as a perturbation problem, we use the notation and formalism of
Cunningham {\em et al.}\cite{CPM} except that we will omit the tildes
over variables; equations from that paper will be cited as ``CPM\@.''
It is important to note that the formulas in that paper, based on the
work of Moncrief\cite{Moncrief}, are gauge invariant, so we need pay no
attention to the coordinate gauge of (\ref{lowest}).

From CPM (II-25),(II-26), we find that the only non-vanishing metric
perturbation functions are $ H_{2}=K=g(r)p(\mu_0) $ where (with the
notation of (\ref{sigmanotat}))
\begin{equation}
g(r)\equiv\frac{M^3}{8R^3\left(1+M/2R\right)}\ \ \ \ \
p(\mu_0)\equiv\frac{\Sigma_{1}(\mu_{0})+
\Sigma_{3}(\mu_{0})}{(\Sigma_{1}(\mu_{0}))^{3}}\,,
\end{equation}
From CPM (II-27),(II-28) we next find
\begin{equation}\label{Qeq}
Q_1=2r\left(1-\frac{2M}{r}\right)^2\left[
\frac{g}{1-2M/r}-\frac{1}{\sqrt{1-2M/r}}
\frac{d}{dr}\left(\frac{rg}{\sqrt{1-2M/r}}\right)\right]+6rg\,,
\end{equation}
and, following CPM (II-31), we define
\begin{equation}\label{psieq}
\psi\equiv\,\sqrt{\frac{4\pi}{5}}\frac{Q_1}{\lambda}\,,
\end{equation}
where$
\lambda\equiv1+\frac{3M}{2r}\,.
$
The function $\psi$ then satisfies the $\ell=2$
Zerilli equation\cite{Zerilli}
\begin{equation}
\frac{\partial^2\psi}{\partial t^2}-
\frac{\partial^2\psi}{\partial r_*^2}
+\left(1-\frac{2M}{r}\right)
\left\{\frac{1}{\lambda^2}\left[\frac{9M^3}{2r^5}-\frac{3M}{r^3}
\left(1-\frac{3M}{r}
\right)
\right]+\frac{6}{r^2\lambda}
\right\}=0\,.\label{zerieq}
\end{equation}
where $r_*\equiv r+2M\ln{\left(\frac{r}{2M}-1\right)}$.  [It is worth
noting that the function $Q_2$ of CPM (II-28), calculated from our
perturbations (\ref{Qeq}), explicitly solves the initial value constraint
$Q_2=0$ as given in CPM (II-29)].

The form of $\psi$ given by (\ref{Qeq})-(\ref{psieq}) is now taken as
initial data along with the initial condition $\partial\psi/\partial
t=0$ at $t=0$.  The problem is greatly simplified by the fact that the
only $\mu_{0}$ dependence is contained in the multiplicative factor
$p(\mu_{0})$. Since the initial data are proportional to $p(\mu_{0})$,
it follows that the evolved waveform $\psi$ is proportional to
$p(\mu_{0})$, and the radiated power and energy are proportional to
$[p(\mu_{0})]^{2}$. It is only necessary, therefore, to do one
computation of the evolved waveform and radiated power. The $\mu_{0}$
dependence is known at the outset.

The initial form of $\psi$ (with $p(\mu_{0})$ set to unity) is shown
in figure 1. We evolved these data numerically with the Zerilli
equation (\ref{zerieq}). The resulting wave form, at $r^*=200$ (we use
units in which $2M=1$), as a function of $t$, is shown in figure 2. It
clearly exhibits quasinormal ringing and power-law tails corresponding
to a quadrupolar perturbation. The values of the quasinormal
frequencies\cite{Leaver} and power-law exponents\cite{GPPI} are in
excellent agreement with theoretical values. From the evolved data we
compute the radiated power, which is given by CPM (III-28),
\begin{equation}
{\rm Power} = {\textstyle {1 \over 348 \pi}} \left|
{\partial \over \partial t} \psi\right|^2 \,.
\end{equation}
If $p(\mu_{0})$ is set to unity, this procedure gives a computed
energy of $3.07\times 10^{-6}$. The result 
energy, as a function of $\mu_{0}$, is therefore
\begin{equation}
{\rm Energy}/2M=3.07\times 10^{-6}p(\mu_{0})^{2}\,.
\end{equation}
This result is displayed in figure 2, where it is compared with the
numerical results reported by Anninos {\em et al.} \cite{Annetal}. It
is intriguing that the remarkable agreement 
extends considerably beyond the small-$\mu_{0}$ region in which  our
approximation is expected to be applicable.

The general method is suited to a fairly wide variety of initial data.
This paper only concentrated on the aesthetically elegant Misner data
as an example. The method can also be applied to initial data which
are known only in numerical form and which represent perturbations of
a black hole. To do this would require considerably more numerical
analysis than for the Misner initial data but, when applicable, would
be much quicker and less expensive than the fully numerical methods
for integrating the nonlinear field equations.

\thanks

This work was initiated as the result of a suggestion by Larry Smarr
at a conference. We wish to thank Larry Smarr, Ed Seidel and Wai-Mo
Suen for discussions. This work was supported by grants
NSF-PHY93-96246, NSF-PHY-92-07225, by research funds of the University
of Utah and Penn State University and Penn State's Minority Faculty
Development Program.

\begin{figure}
\caption{The function $\psi$ of the Cunningham-Price-Moncrief
perturbation scheme for Misner's initial data. The values shown are
for $p(\mu_{0})=1$, and for units in which $2M=1$. For the $r^*$ 
coordinate the horizon is at $r^*=-\infty$ and $r^* \sim r$ for large
positive values. }
\end{figure}
\begin{figure}
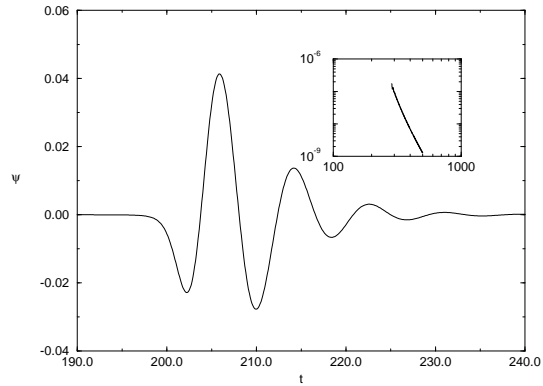

\caption{Time evolution of the Misner initial data (with
$p(\mu_{0})=1, \ 2M=1$), from the point of view
of an observer fixed at $r^*=200$. We see the appearance of
quasinormal ringing with the predicted period of 8.4. In the inset we
display in a log-log plot the late time behavior of the field, which
clearly exhibits a power-law tail form with exponent $-6$ as predicted
by theory.}
\end{figure}
\begin{figure}
\caption{The solid curve is the prediction for the radiated energy, 
as a function of 
$\mu_{0}$, based on linearized perturbation theory. The black dots
correspond to the values of numerical relativity results reported by
Anninos {\em et al.} [5]. }
\end{figure}


\begin{references}


\bibitem{Ab} A. A. Abramovici {\em et al.}, Science {\bf 256}, 325
(1992), and references therein.
\bibitem{Cuetal} C. Cutler {\em et al.}, Phys. Rev. Lett. 
{\bf 70}, 2984 (1993). 
\bibitem{SmEp}
L. L. Smarr, Ph.D. dissertation, University of
Texas at Austin, unpublished (1975);
K. R. Eppley, Ph.D. dissertation, Princeton University,
unpublished (1977).
\bibitem{smarrsources}
L. L. Smarr, in {\it Sources of
Gravitational Radiation,} ed. L. L. Smarr
(Cambridge University Press, Cambridge, 1979).
\bibitem{Annetal} P. Anninos, D. Hobill, E. Seidel, L. Smarr, W.-M.
Suen, Phys. Rev. Lett. {\bf 71}, 2851 (1993).
\bibitem{Misner}C. Misner, Phys. Rev. {\bf 118}, 1110 (1960).
\bibitem{CPM}C. Cunningham, R. Price, V. Moncrief, Ap. J. {\bf 230},
870 (1979).
\bibitem{Moncrief} V. Moncrief, Ann. Phys. (NY), {\bf 88}, 323 (1974).
\bibitem{Zerilli}F. Zerilli, Phys. Rev. Lett. {\bf 24}, 737 (1970).
\bibitem{Leaver}E. W. Leaver, Proc.~R. Soc.~London {\bf A402}, 285 (1985).
\bibitem{GPPI}C. Gundlach, R. H. Price and J. Pullin, Phys. Rev. {\bf D49}, 
883 (1994).
\end{references}
\end{document}